\documentclass[5p,number,times]{elsarticle}

\usepackage{graphicx}
\usepackage[figuresright]{rotating}
\usepackage{amsmath}
\usepackage{amssymb}
\usepackage{latexsym}
\usepackage{multirow}
\usepackage{topcapt}
\usepackage{url,hyperref}

\title{Simulation of neutrons produced by high-energy muons underground}
\author[lip]{A.~Lindote\corref{cor1}}
\ead{alex@lipc.fis.uc.pt}
\author[icl,ral]{H.~M.~Ara\'{u}jo}
\author[shef]{V.~A.~Kudryavtsev}
\author[shef]{M.~Robinson}
\address[lip]{LIP--Coimbra and Department of Physics of the University of Coimbra, Portugal}
\address[icl]{Blackett Laboratory, Imperial College London, UK}
\address[ral]{Particle Physics Department, STFC Rutherford Appleton Laboratory, UK}
\address[shef]{Department of Physics and Astronomy, University of Sheffield, UK}

\cortext[cor1]{Corresponding author: \href{mailto:alex@lipc.fis.uc.pt}{alex@lipc.fis.uc.pt}}

\begin{document}

\begin{abstract}
This article describes the Monte Carlo simulation used to interpret the measurement of the muon-induced neutron flux in the Boulby Underground Laboratory (North Yorkshire, UK), recently performed using a large scintillator veto deployed around the ZEPLIN-II WIMP detector. Version 8.2 of the GEANT4 toolkit was used after relevant benchmarking and validation of neutron production models. In the direct comparison between Monte Carlo and experimental data, we find that the simulation produces a 1.8 times higher neutron rate, which we interpret as over-production in lead by GEANT4. The dominance of this material in neutron production allows us to estimate the absolute neutron yield in lead as~(1.31~$\pm$~0.06)~$\times$~10$^{-3}$~neutrons/muon/(g/cm$^2$) for a mean muon energy of 260~GeV. Simulated nuclear recoils due to muon-induced neutrons in the ZEPLIN-II target volume ($\sim$1 year exposure) showed that, although a small rate of events is expected from this source of background in the energy range of interest for dark matter searches, no event survives an anti-coincidence cut with the veto.
\vspace{1pc}
\end{abstract}

\maketitle

\section {Introduction}
\label{intro}

Neutrons constitute a very important background for experiments looking for rare events in deep underground laboratories. The vast majority is produced by natural radioactivity in the rock or the materials used in the detectors: traces of U/Th emit $\alpha$ particles which may then interact with light nuclei (Z$\lesssim$30) to produce neutrons; another contribution comes from spontaneous fission of heavy elements (mainly $^{238}$U). The energy range of these neutrons is restricted to a few MeV, and it is therefore possible to shield detectors from rock neutrons using H-rich materials, which moderate and capture them. Radioactivity from detector parts can in turn be minimised by an appropriate choice of building materials and selection of batches. Moreover, these local sources of radioactivity can be assessed with dedicated measurements (see, for example, Ref. \cite{nuboulby} for the case of the Boulby Underground Laboratory). Detailed Monte Carlo (MC) simulations can then use this information to optimise the geometry of the passive shielding and to predict this contribution to the total background of the experiment, thus reducing its systematic uncertainties.

A much smaller contribution comes from muon interactions in the rock and surrounding materials, which produce fast neutrons with energies up to the GeV scale. These high energy neutrons can easily penetrate through passive shielding (which can in fact act as a target for their production) and interact in the detectors. In dark matter searches, elastic scattering of high energy neutrons produces nuclear recoils within the expected energy range of interactions from Weakly Interacting Massive Particles (WIMPs). In double beta decay experiments they contribute to the $\gamma$-ray background via inelastic scattering and capture; they also mimic neutrino detection in scintillators via inverse beta decay. Active vetoes are thus necessary to help remove this background, by looking at coincidences with high energy deposits from the primary muon or the associated cascade. In spite of this, these neutrons can travel far from the original muon track, interacting in the detector while leaving no telltale signature in the anti-coincidence system or the detector itself. We show in Section \ref{results} that the rate expected from such events in the energy range of interest is quite low for the current generation of dark matter experiments based on liquid rare gases, which have only a few kilograms of active material. But with projects under way to build detectors with hundreds or thousands of kilograms, the precise knowledge of this neutron flux and the ability to model it correctly becomes paramount for the design of both the detectors and the associated anti-coincidence systems. Moreover, current detectors would also benefit from this, as it would allow for a reduction in the systematic errors of the expected background.

Measuring the flux of muon-induced neutrons is not trivial, as the requirements are almost those of a low-background experiment: the detector must be sensitive to neutrons and have a large mass, it must be placed in an underground laboratory, and it must be stable throughout the duration of the measurement, which is typically a few months long. Several experiments have tried to measure this flux in the last decades, using either accelerators on the surface \cite{na55} or dedicated experiments in underground laboratories \cite{bezrukov73,asd,lsd,lvd,menghetti04,paloverde,chen,gorshkov74,bergamasco73}, and several groups have projects for new measurements (see, for example, \cite{akerib,kozlov}). But these results are often mutually inconsistent (see \cite{vak03,ha05,wang01} for a more detailed discussion of available data). Moreover, at the time when these experiments were performed detailed MC simulations were not possible due to the lack of computational power and adequate theoretical models.

Powerful simulation packages based on advanced theoretical models, such as GEANT4 \cite{geant4} and FLUKA \cite{fluka,fluka08-1,fluka08-2}, are now available, and although they had some success in explaining the neutron yield in low-A materials, both fail to explain older measurements of neutron production in lead, as was investigated in \cite{ha05}. Extracting the neutron yield from available data is not straightforward, as it requires a detailed description of all physical processes involved and of the geometrical setup used in the experiment. Moreover, such simulations are not trivial either: a good knowledge of the muon flux and energy spectrum for the particular underground site is essential, as they determine the neutron energy spectrum and yield; the statistical efficiency is very low, i.e.~a very large number of muons has to be simulated in order to produce a statistically significant result; and it is all too easy to bias the final result while trying to improve this efficiency.

Recent works have used these simulation packages to try to explain published data \cite{ha05,marino07} or to predict the muon-induced neutron background in underground experiments and optimise the shielding and veto systems (e.g. \cite{bauer06,markusPhD}). 

New measurements of total neutron yield in different materials, the energy spectrum and lateral distribution regarding the primary muon track are therefore urgent, but should be supported by detailed MC simulations, preferably using multi-pur\-po\-se packages (such as GEANT4 and FLUKA) so that their results can be applied to the design of other experiments.

A new measurement of muon-induced neutrons has recently been performed in the Boulby Underground Laboratory (Boulby mine, North Yorkshire, UK) \cite{vitaly08}. A large mass of lead was used as a target that completely surrounded the experimental apparatus. The prime goal of this paper is to describe the detailed Monte Carlo simulations carried out to predict the rate of events due to muon-induced neutrons in a scintillator used in the aforementioned experiment (Section \ref{simulation}). Accurate Monte Carlo of all physical processes and of the whole setup including trigger conditions and selection cuts, and the comparison of the simulation results with the measurements allowed us, for the first time, to test theoretical models directly and to make a robust prediction of the neutron yield in lead (Section \ref{results}). The result for lead is particularly important since i) lead is used as a shielding against gamma-rays in high-sensitivity experiments for rare event searches, such as direct dark matter, double-beta decay and neutrino experiments; ii) previous measurements of muon-induced neutron flux in lead reported a higher yield compared to simulations (carried out, however, by different teams not connected to the original experiments).

We have also performed detailed comparison of our Monte Carlo with previous simulations using GEANT4 and FLUKA, and investigated the dependence of the neutron yield on the models used in simulations (Section \ref{models}). We show (Section \ref{models}) that our simulations agree within a factor of 2 with most available data on neutron yields in light targets but cannot explain some of the previously reported measurements, in particular in heavy targets such as lead. This is probably due to the fact that our models were too simplified and did not take into account all details of the experimental setup (detector configuration, trigger, cuts, etc.) that were not publicly available.

The procedure for muon-induced neutron modelling descri\-bed in this paper can also be used in similar simulations for any other underground experiment for rare event searches.

\section {Model validation}
\label{models}

Processes responsible for neutron production by muons can be categorised in four main classes: \emph{i)} muon capture and \emph{ii)} direct muon spallation result from direct interactions of muons with nuclei. In thick targets showers initiated by a muon-indu\-ced spallation are the dominant source of neutrons: \emph{iii)} photo- and lepto- production, mainly in electromagnetic cascades, \emph{iv)} ha\-dro-production, mainly in hadronic cascades.

Negative muon capture is only dominant for shallow depths ($\lesssim$100 m w.e.), where abundant stopping muons can be captured resulting in highly excited isotopes which then emit one or more neutrons. In direct muon spallation the muon-nucleus interaction may be described through the exchange of a virtual photon, resulting in nuclear disintegration. This treatment allows the use of measured $\gamma$-N cross sections (real photons) to model this process, but breaks down at low energies, when the virtuality of the photon becomes comparable to the muon energy \cite{wang01}.

Many particles and mechanisms are involved in neutron production within showers, and so it is not surprising that many physics models are required to describe them over different energy ranges. In addition, GEANT4 may offer alternative models to treat each physical process, which can be selected according to the particular requirements of each simulation (e.g. the trade-off between computational efficiency and accuracy is often considered). Some of the models available for describing interactions of muons and their associated showers were tested previously \cite{ha05} using version 6.2 of the toolkit against other simulation results; in this work we use a similar set of models with version 8.2:
\begin{itemize}
\item muon-induced spallation: available above 1 GeV muon energy; the resulting final states are obtained  from parameterised hadronic models.
\item photo-production ($\gamma$ inelastic scattering): final states are generated by a chiral-invariant phase-space (CHIPS) decay model below 3 GeV photon energy, while a theoretical quark-gluon string (QGS) model is used at higher energies.
\item hadronic interactions of nucleons, pions and kaons: a QGS model is used for high energies  and an intra-nuclear binary cascade (BiC) for low energies. An older parameterised model (LEP) can be used to cover the intermediate range. This is discussed in more detail further ahead.
\item de-excitation of the residual nucleus: $\gamma$ and fragment evaporation, fission, Fermi break-up and multi-fragmentation (for highly excited nuclei).
\item neutron interactions are treated using data-driven models below 19 MeV.
\end{itemize}

For electromagnetic interactions the so-called `Low Energy' package was used with production thresholds of a few tens of keV for $\gamma$-rays and $\sim$1 MeV for e$^{-}$ and e$^{+}$ in all materials. Such low thresholds are justified by the importance of the photo-nuclear process as a significant source of neutrons.

The importance of the energy scale at which the changeover between the BiC and the QGS hadronic models occurs was studied. The former aims to describe interactions in the low energy range (Ref. \cite{g4models} recomends its usage below 3~GeV for $p$ and $n$, and 1.5 GeV for pions, pointing out that it should work reasonably well up to 10~GeV for $p,n$). At very low energies (below $\sim$70~MeV, but the real threshold depends on atomic mass, $A$) it reverts to a `precompound' model, which handles the nuclear de-excitation in the pre-equilibrium stage. On the other hand, QGS is targeted at high energies (above 20~GeV) and depends on other models to fragment and de-excite the damaged nucleus after the initial interaction. This may involve either the precompound model (this association is usually known as QGSP) or the CHIPS model (QGSC). There is currently no theoretical model available to cover the energy region between the BiC and QGS models; the solution commonly adopted is a low-energy parameterisation model (LEP), but this is not kinematically correct for a single interaction \cite{g4models}.

We analysed different possibilities to bridge this energy range. Firstly the LEP model was avoided altogether for $p$ and $n$, using changeover energies of 6~GeV (first test) and of 10~GeV (second test) between the BiC and QGSP models. A third test relied on LEP for $p$ and $n$ between 10 and 20~GeV (this is very similar to the approach used in the QGSP\_BIC\_HP reference physics list provided by GEANT4 \cite{g4ref_lists}). As for pions we never extended the recommended range of the BiC model (1.5 GeV), and thus had to use LEP to bridge to the QGSP minimum energy (note that this is unlike the QGSP\_BIC\_HP reference physics list, where the LEP model is used down to low energies). For all other hadrons, low- and high-energy parameterised models were used.

These approaches were tested in various materials for different muon energies. We found that the resulting neutron yields always agreed within 10\%, which is a smaller discrepancy than that observed between versions 6.2 and 8.2 of the toolkit. \linebreak GEANT4 offers an alternative model to BiC for the treatment of low energy hadronic reactions, the Bertini Cascade \cite{g4models}: this is similarly found not to alter the total yields noticeably \cite{luciano08}. We thus opted for a direct changeover at 6~GeV between BiC and QGSP for $p$ and $n$, as it enables a more direct comparison with earlier work \cite{ha05}.

\subsection{Total yield in different materials}
 \label{neutron_yield}

We began by benchmarking the neutron yield simulations reported in Ref.~\cite{ha05} for different materials using a similar setup. This consisted of a beam of mono-energetic $\mu^-$ incident at the centre of a slab of material with thickness 3200 g/cm$^2$; only neutrons produced in the central half-length of the slab were counted to avoid edge effects. Treatment of secondary neutrons produced in inelastic scattering was also considered to prevent double counting. 

\begin{figure}[htb]
\begin{center}
   \includegraphics[width=1.0\columnwidth]{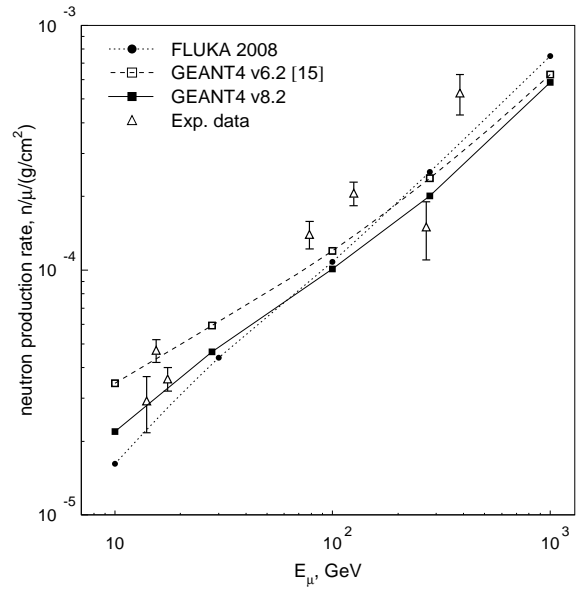}
    \caption{{Variation of the neutron yield (per unit muon track length) with the initial muon energy for C$_{n}$H$_{2n}$ scintillator. Experimental data are taken from measurements at various depths between 20 and 5200 m w.e. using the mean muon energy for the respective depth. Lines connecting the markers are to guide the eye only.}}
  \label{fig-c10h20yield}
\end{center}
\end{figure}

As most of the neutron yield data available is for organic scintillators, we started by studying a C$_{10}$H$_{20}$ hydrocarbon with density $\rho=0.8$ g/cm$^3$. Nevertheless, these results are extensible to generic hydrocarbons C$_{n}$H$_{2n}$, which can also represent materials usually used as passive neutron shielding. Figure \ref{fig-c10h20yield} shows the neutron yield as a function of the incident muon energy for this material. Statistical uncertainties are comparable to the size of the markers. Also shown for comparison are the yields obtained with GEANT4 version 6.2 (as reported in Ref. \cite{ha05}) and FLUKA-2008. The neutron yields obtained with FLUKA-2008 are in good agreement to what was reported in Refs. \cite{vak03,wang01} using FLUKA-1999, but lower (especially at low energies) than the FLUKA-2003 results reported in Ref. \cite{ha05}. It is clear that version 8.2 of GEANT4, while still consistent with the experimental measurements, produces systematically fewer neutrons than version 6.2. At lower energies ($\lesssim100$~GeV) the yield is closer to that predicted by FLUKA, while the agreement with version 6.2 of GEANT4 is better at higher energies.

\begin{figure}[htb]
\begin{center}
   \includegraphics[width=1.0\columnwidth]{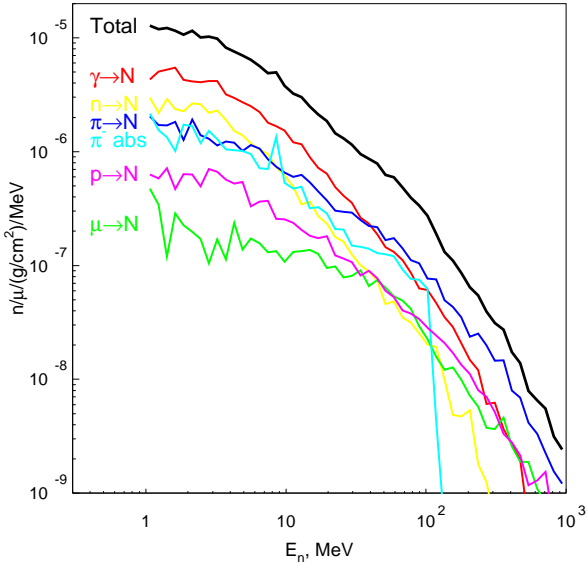}
    \caption{{Differential energy spectrum of neutrons produced in C$_{n}$H$_{2n}$ by 280~GeV muons (in thick black). Also shown are the contributions of the most important individual processes: photonuclear interaction of $\gamma$-rays ($\gamma \rightarrow N$), neutron inelastic scattering ($n \rightarrow N$), pion spallation ($\pi \rightarrow N$) and absorption ($\pi^- abs$), and proton ($p \rightarrow N$) and muon ($\mu \rightarrow N$) spallation.}}
  \label{fig-c10h20_280_proc}
\end{center}
\end{figure}

Figure \ref{fig-c10h20_280_proc} shows the differential energy spectrum of neutrons produced by 280~GeV muons in this material as well as the contributions from the most important individual processes. The photonuclear interaction of $\gamma$-rays dominates for low energy neutrons ($\lesssim$40~MeV), while pion spallation is the biggest contributor above this energy. The pion absorption process shows the expected cut-off right below the pion rest mass subtracted by the nucleon binding energy, but unlike what has been reported in Ref. \cite{markusPhD}, where the classical GEANT4 nuclear capture model was used, does not exhibit a distinct peak before this cut-off. In this work we used a new CHIPS-based model, which produces a smoother spectrum with more low energy neutrons. Muon spallation, responsible for initiating the electromagnetic and hadronic showers, clearly contributes very little to direct neutron production for muon energies above 100~GeV. Version 6.2 of GEANT4 yields very similar results for all processes, with the notable exception of the photonuclear interaction for which it produces more low energy neutrons ($\lesssim$20~MeV).

\begin{figure}[htb]
\begin{center}
   \includegraphics[width=1.0\columnwidth]{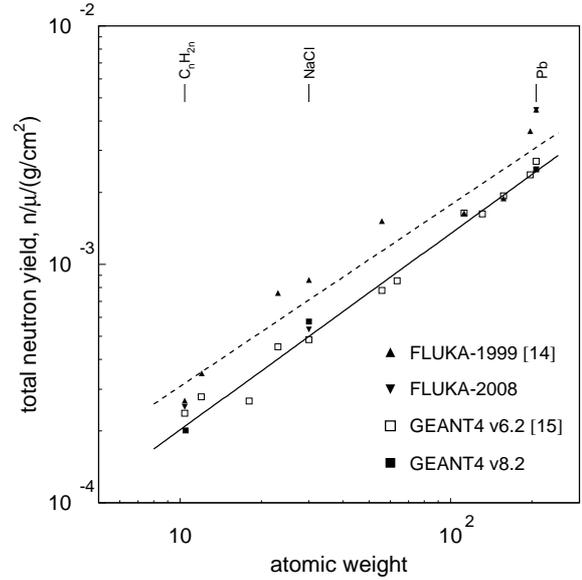}
    \caption{{Neutron yield as a function of the average atomic weight for incident muons of 280 GeV. Shown for comparison are the results obtained with FLUKA-1999 in \cite{vak03}, GEANT4 6.2 \cite{ha05} (and the respective power-law parameterisations) and FLUKA-2008.}}
  \label{fig-matsyield}
\end{center}
\end{figure}

We also tested other materials of relevance for low background experiments, namely NaCl (which dominates the composition of Boulby rock) and Pb (usually used as passive $\gamma$-ray shielding). For this study we used 280 GeV muons, close to the mean muon energy at Boulby ($\sim$260~GeV), and kept the selection rules mentioned previously. Figure \ref{fig-matsyield} shows the total neutron yield as a function of atomic weight (average atomic weight for compounds). Results obtained with FLUKA-1999 \cite{vak03}, GEANT4 6.2 \cite{ha05} and FLUKA-2008 are shown for comparison. For the latter only C$_n$H$_{2n}$, NaCl and Pb were tested, while for the former results for several materials are presented along with the respective fits to the power law $R=bA^\beta$ ($R$ is the neutron rate and $A$ is the atomic weight). GEANT4 8.2 yields are consistent with those obtained with version 6.2, while FLUKA-1999 predicts consistently higher rates: $\sim$30\% higher in C$_{n}$H$_{2n}$, $\sim$50\% in NaCl and $\sim$80\% in Pb. Yields from the new FLUKA-2008 are very similar to the ones reported for FLUKA-1999 in Ref. \cite{vak03} for both C$_n$H$_{2n}$ and Pb, but much lower for NaCl, agreeing with GEANT4 predictions in this case. 

\begin{figure}[htb]
\begin{center}
   \includegraphics[width=1.0\columnwidth]{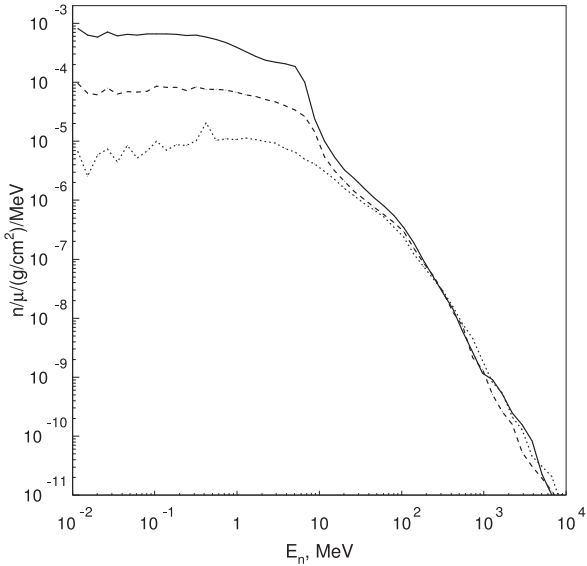}
    \caption{{Differential energy spectra for neutrons produced in Pb (thick lines), NaCl (dashed lines) and C$_{n}$H$_{2n}$ (dotted lines) for 280~GeV incident muons (except for Pb, in which case 260~GeV muons were used). Mean energies of these distributions are 8.8~MeV, 23.4~MeV and 65.3~MeV for Pb, NaCl and C$_{n}$H$_{2n}$ respectively.}}
  \label{fig-mats-spectra}
\end{center}
\end{figure}

Figure \ref{fig-mats-spectra} shows the spectra for neutrons produced in each of these materials using incident muons of 280~GeV (for NaCl and C$_{n}$H$_{2n}$) and 260~GeV (for Pb). It is clear that the increase in neutron yield seen in materials with large atomic number comes mainly from low energy neutrons ($\lesssim$20~MeV). These spectra can be compared with the spectra from rock radioactivity (see Ref. \cite{nuboulby} for the case of Boulby): while neutrons from the rock are restricted to a few MeV, muon induced neutrons extend to tens and even hundreds of MeV.

In conclusion, the neutron yields from GEANT4 have not changed significantly from version 6.2 to 8.2 for the materials and energies of interest for our experimental setup. Differences to the FLUKA results are still present and in some cases even accentuated, with the latter producing nearly twice as many neutrons in lead. As was discussed in Ref. \cite{ha05} this difference in the neutron yield may originate from an over-production in hadronic cascades by FLUKA: as the emission of fast nuclear fragments from highly excited nuclei is not modelled, more energy is available for neutron evaporation.

\subsection{Muon-induced spallation}
 \label{mu_spallation}
 
As mentioned previously, the muon-nucleus interaction can be modelled by the exchange of a virtual photon, which allows the use of cross-sections parameterised for the case of the real photonuclear interaction. For thick targets this is not a dominant process for neutron production, and its importance decreases with both energy and atomic weight \cite{ha05,wang01}. Nevertheless we discuss it here in more detail in order to test a new implementation based on the CHIPS concept, which has become available in recent versions of GEANT4.
 
CHIPS (see \cite{g4physics,chips} and references therein) was first available in this toolkit as a nuclear de-excitation model, meant to be used in conjunction with other models (such as the QGS) which handled the initial interaction. Recently, new CHIPS-based implementations have become available which include the cross-sections of the primary interaction and simulate the entire reaction for a number of projectiles including muons.
 
\begin{figure*}[htb]
\begin{center}
   \includegraphics[width=1.0\columnwidth]{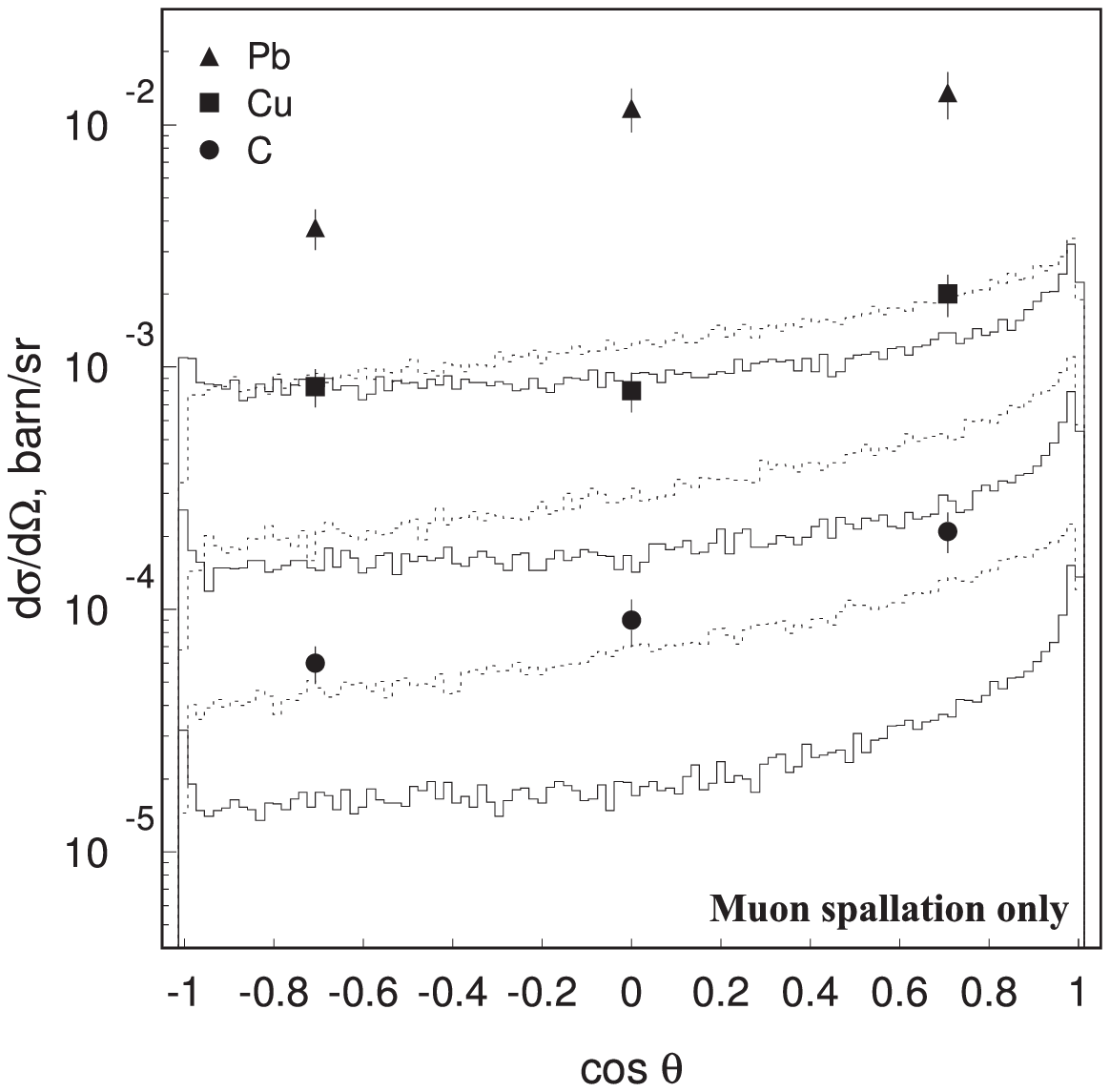}
   \includegraphics[width=1.0\columnwidth]{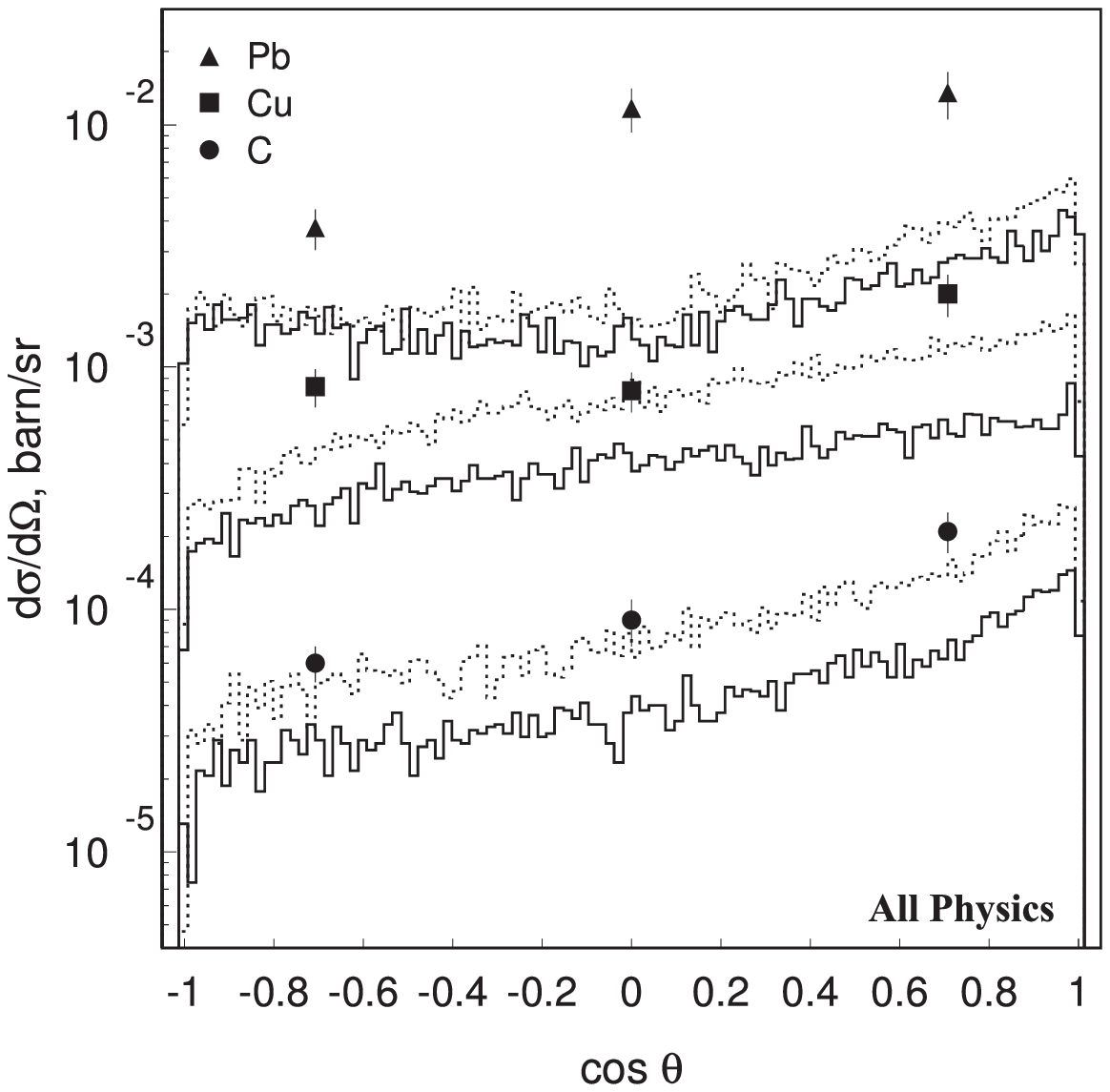}
    \caption{{Differential cross-section of neutron production in (curves from bottom to top) graphite, copper and lead by 190 GeV muons (for a 10 MeV threshold). Plot on the left shows GEANT4 simulations considering only the direct muon spallation interaction -- cases $a)$ and $c)$ described in the text -- while the one on the right includes the complete physics lists -- cases $b)$ and $d)$. Solid lines show standard process results; dotted lines the new CHIPS-based processes; markers show the NA55 results.}}
  \label{fig-na55-yield}
\end{center}
\end{figure*}

As reported in Refs.~\cite{ha05,marino07} the older GEANT4 model for the simulation of muon-induced spallation (MuNuclear) failed to describe the results obtained by the CERN NA55 experiment \cite{na55}, which claimed to have measured fast neutron production in graphite, copper and lead using a beam of 190 GeV muons. Neutron detectors with a reported threshold of $\sim$10 MeV were placed at 45$\rm{^o}$, 90$\rm{^o}$ and 135$\rm{^o}$ regarding the initial muon beam. However, the thin target assumption in this experiment has been questioned \cite{ha05}.  We investigated whether the newly available models could explain the discrepancies found previously. Four simulations were conducted: $a)$ using only the previous GEANT4 muon spallation process; $b)$ using the complete set of physics processes described previously; $c)$ using only the new CHIPS model for the simulation of muon photonuclear interactions; and $d)$ using the complete set of physics processes, but selecting CHIPS versions whenever available (na\-me\-ly for all lepto-nuclear interactions, capture of negatively charged particles at rest, and nuclear de-excitation). Note that $a)$ and $c)$ should produce results similar to $b)$ and $d)$, respectively, if the thin target assumption is correct. 

Figure \ref{fig-na55-yield} shows the differential cross-sections obtained from considering only the primary muon interaction (left) or all the physics processes (right). Results with the standard physics processes ($a)$ and $b)$) reproduce those obtained with older versions \cite{ha05,marino07}, and fail to explain the experimental results. The new CHIPS model for muon spallation produces a dramatic increase for graphite, showing a reasonable agreement with the data; using the complete set of physics processes does not chan\-ge this result. As for copper, the primary process by itself already produces more neutrons than the older one, but in this case it is only the use of the complete physics list that brings the simulation result closer to the measured yields, showing that this could not be considered as a thin target. 

The very large discrepancy observed for lead remains unexplained, with the CHIPS-based models yielding approximately the same result as the older versions, about one order of magnitude lower than the measurement. Considering a lower energy threshold for the neutron detectors helps to reduce this difference, but a value as low as 3 MeV is required for a reasonable agreement. In any case the shape of the distribution still disagrees with the measurements.

\begin{figure*}[hp]
\begin{center}
   \includegraphics[width=1.9\columnwidth]{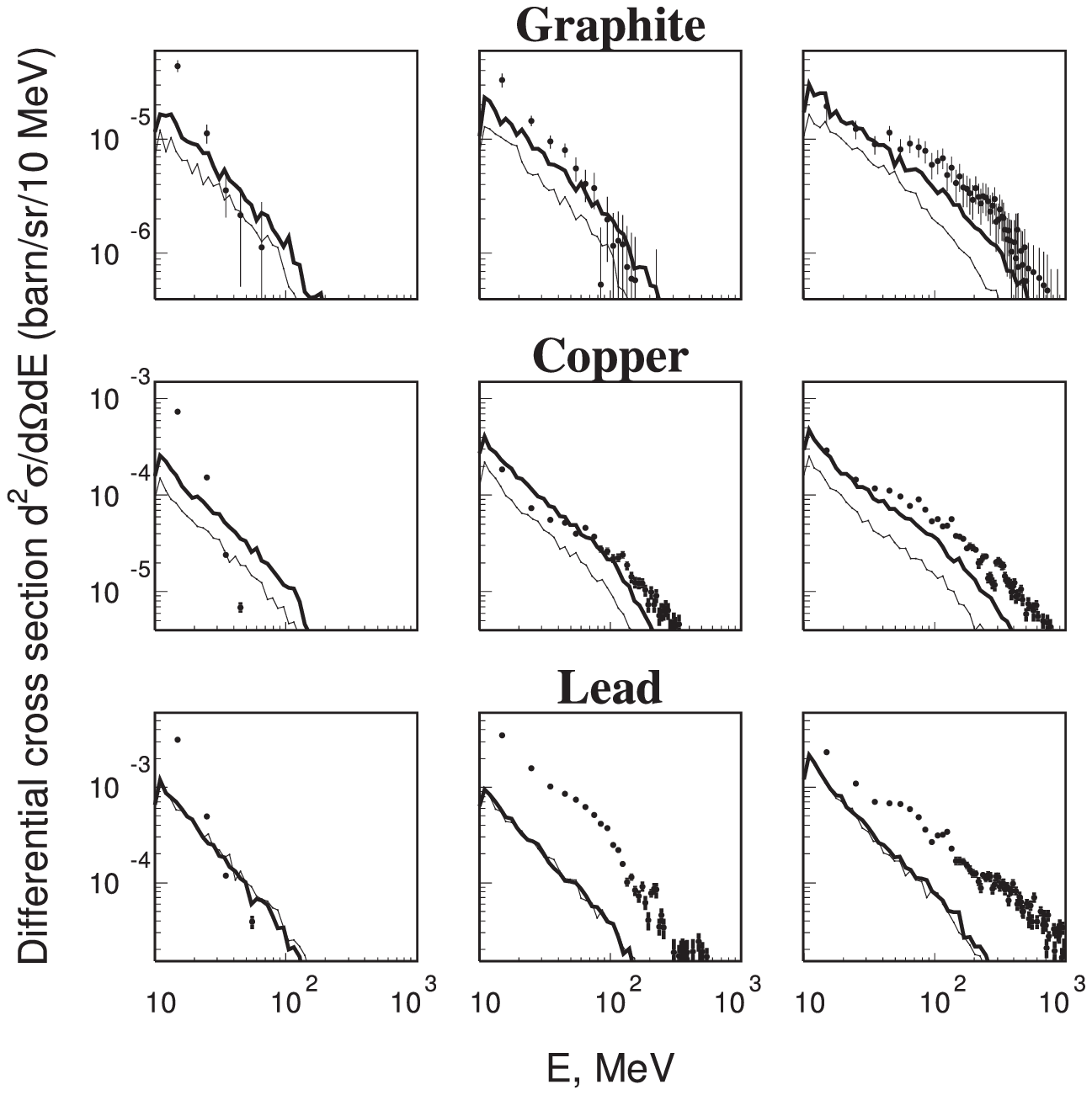}
    \caption{{Energy spectra of neutrons produced in (top to bottom) graphite, copper and lead for 135$\rm{^o}$, 90$\rm{^o}$ and 45$\rm{^o}$ (1st, 2nd and 3rd columns, respectively). Thick lines show simulation results using CHIPS models -- case $d)$ described in the text; thinner lines using the models described in section \ref{models} -- case $b)$. Data points were taken from the NA55 publication \cite{na55}.}}
  \label{fig-na55-energy}
\end{center}
\end{figure*}

Figure \ref{fig-na55-energy} compares the neutron energy spectra using physics lists with complete sets of processes (cases $b)$ and $d)$) for each of these materials. Results from both simulations are more similar in the case of lead, but as already indicated from Figure \ref{fig-na55-yield} completely fail in reproducing the experimental results for all three angles. CHIPS produces a better agreement for graphite, but is still lower for 45$\rm{^o}$ at high energies. The differences between simulation and data are not as evident in copper as in lead, but both models still clearly fail in describing their behaviour.

From the analysis of Figures \ref{fig-na55-yield} and \ref{fig-na55-energy} it is clear that none of the tested models is able to describe the whole set of results from the NA55 experiment. However, this disagreement should be taken with caution: it is difficult to do such simulations retrospectively based on the little information available in the literature. This reafirms the need for new experiments accompanied by dedicated and detailed MC simulations of the experimental setup and selection criteria.
 
\subsection{Muon transport}
 \label{mu_transport}
 
We also tested muon transport in GEANT4 against FLUKA-2007 and the muon propagation code MUSIC \cite{music}. For the case of GEANT4 the two models described in \ref{mu_spallation} to handle direct muon spallation (MuNuclear and CHIPS) were used, but no other change was done to the physics list. 

\begin{figure}[htb]
\begin{center}
   \includegraphics[width=1.0\columnwidth]{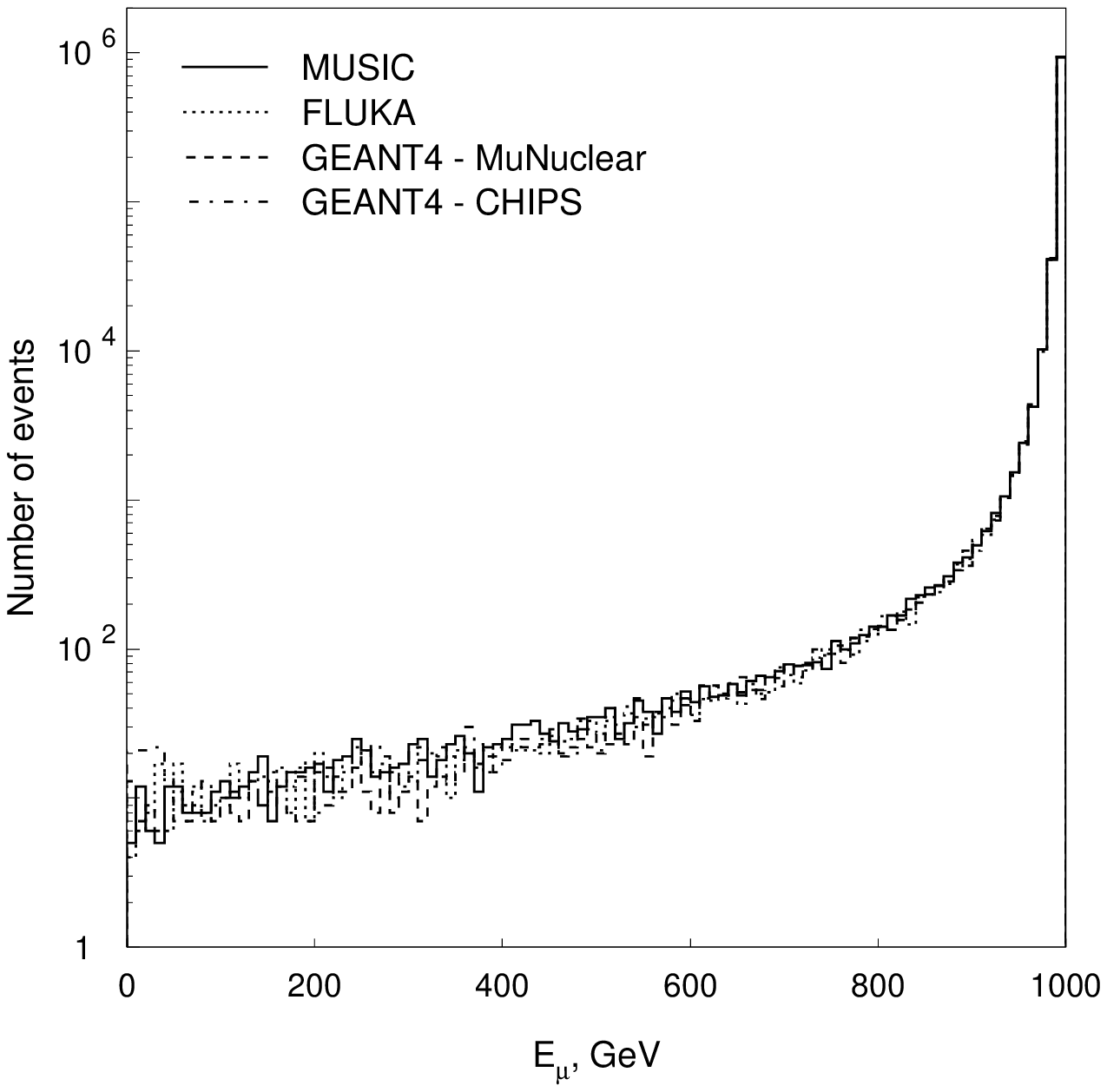}
    \caption{{Energy spectrum of 1000 GeV muons after crossing a 10~m thick slab of water.}}
  \label{fig-transport_10mwe}
\end{center}
\end{figure}

First, one million muons with 100 GeV and 1000 GeV were propagated through slabs of different materials with a thickness equivalent to 10 m w.e. Figure \ref{fig-transport_10mwe} shows the energy distributions of 1000~GeV muons upon exiting a slab of water. Both GEANT4 models agree well with MUSIC and FLUKA, with the only exception of a factor of 2 enhancement in the number of muons below 50 GeV (5\% of the initial energy) for the CHIPS model. We would like to point out, however, that, since electromagnetic processes, such as ionisation, bremsstrahlung and pair production dominate in muon energy losses, the effect of muon inelastic scattering (or hadroproduction) can hardly be seen with this approach.

\begin{figure}[htb]
\begin{center}
   \includegraphics[width=1.0\columnwidth]{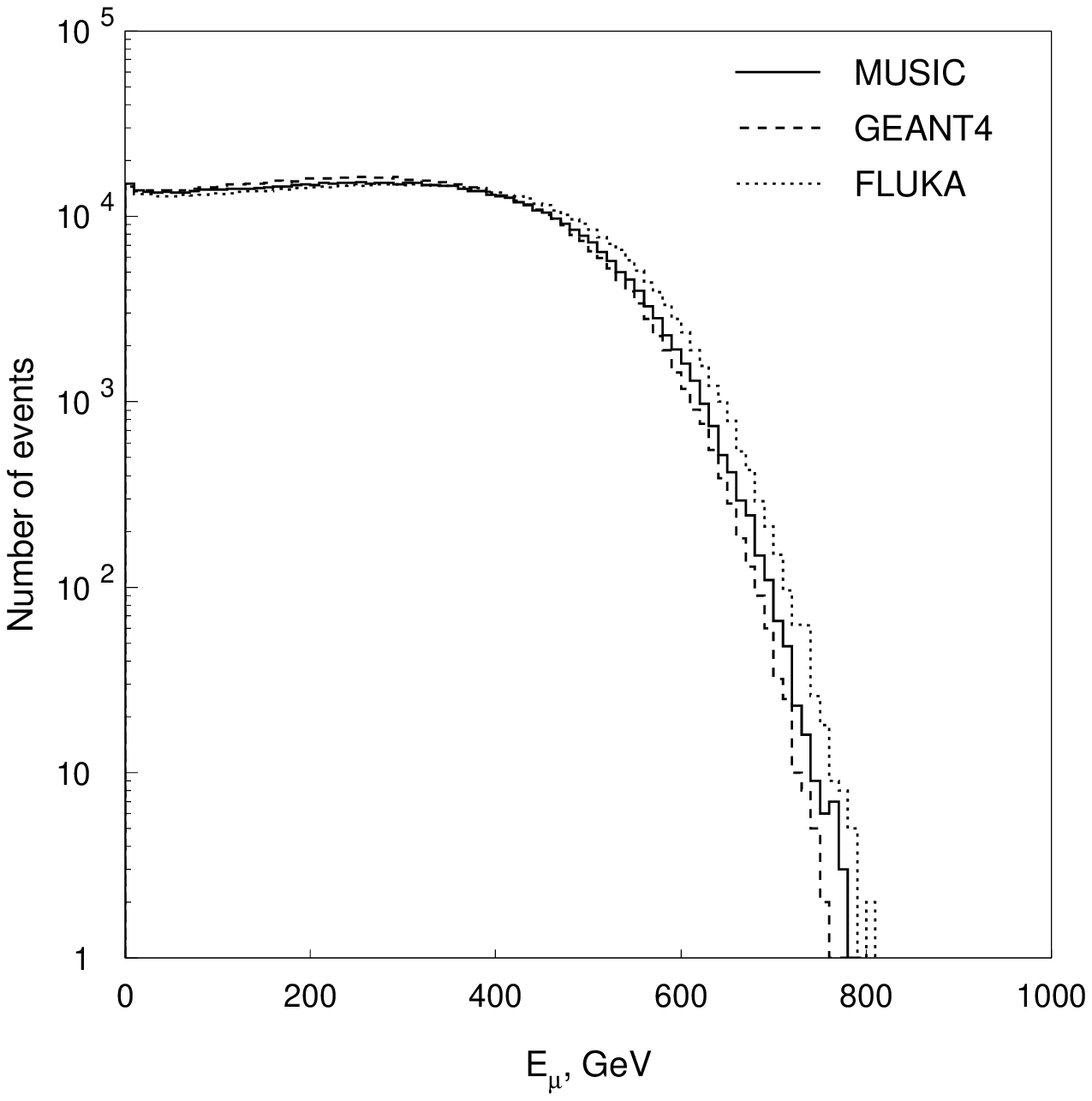}
    \caption{{Energy spectrum of 2 TeV muons after crossing 3 km w.e. of standard rock (Z=11, A=22, $\rho$=2.65 g/cm$^3$). Mean muon energies obtained from MUSIC, GEANT4 and FLUKA are 261, 256 and 273 GeV, respectively, with survival probabilities of 0.730, 0.750 and 0.741.}}
  \label{fig-transport_3km_rock}
\end{center}
\end{figure}

We have also checked the reliability of GEANT4 to propagate muons through a large thickness of matter. This is relevant for the calculation of muon fluxes and spectra at large depths underground, and to predict muon event rates from high-energy neutrino interactions. One million muons with 2 TeV energy have been transported through 3 km w.e. of standard rock (Z~=~11, A~=~22, density~=~2.65 g/cm$^3$). Figure \ref{fig-transport_3km_rock} shows the resulting energy distributions from the three codes: GEANT4, MUSIC and FLUKA. The two GEANT4 models predict very similar energy spectra and only the results using MuNuclear are shown in the figure. Here again the differences between the two models for muon inelastic scattering are hidden in the much larger energy losses due to electromagnetic interactions. The probability for a 2 TeV muon to survive after 3 km w.e. in standard rock is: 0.730 (MUSIC), 0.750 (GEANT4) and 0.741 (FLUKA), with a typical statistical error of 0.1\%. Mean muon energy after propagation is: 261 GeV (MUSIC), 256 GeV (GEANT4), 273 GeV (FLUKA), with a statistical uncertainty of less than 1 GeV.

\begin{figure}[htb]
\begin{center}
   \includegraphics[width=1.0\columnwidth]{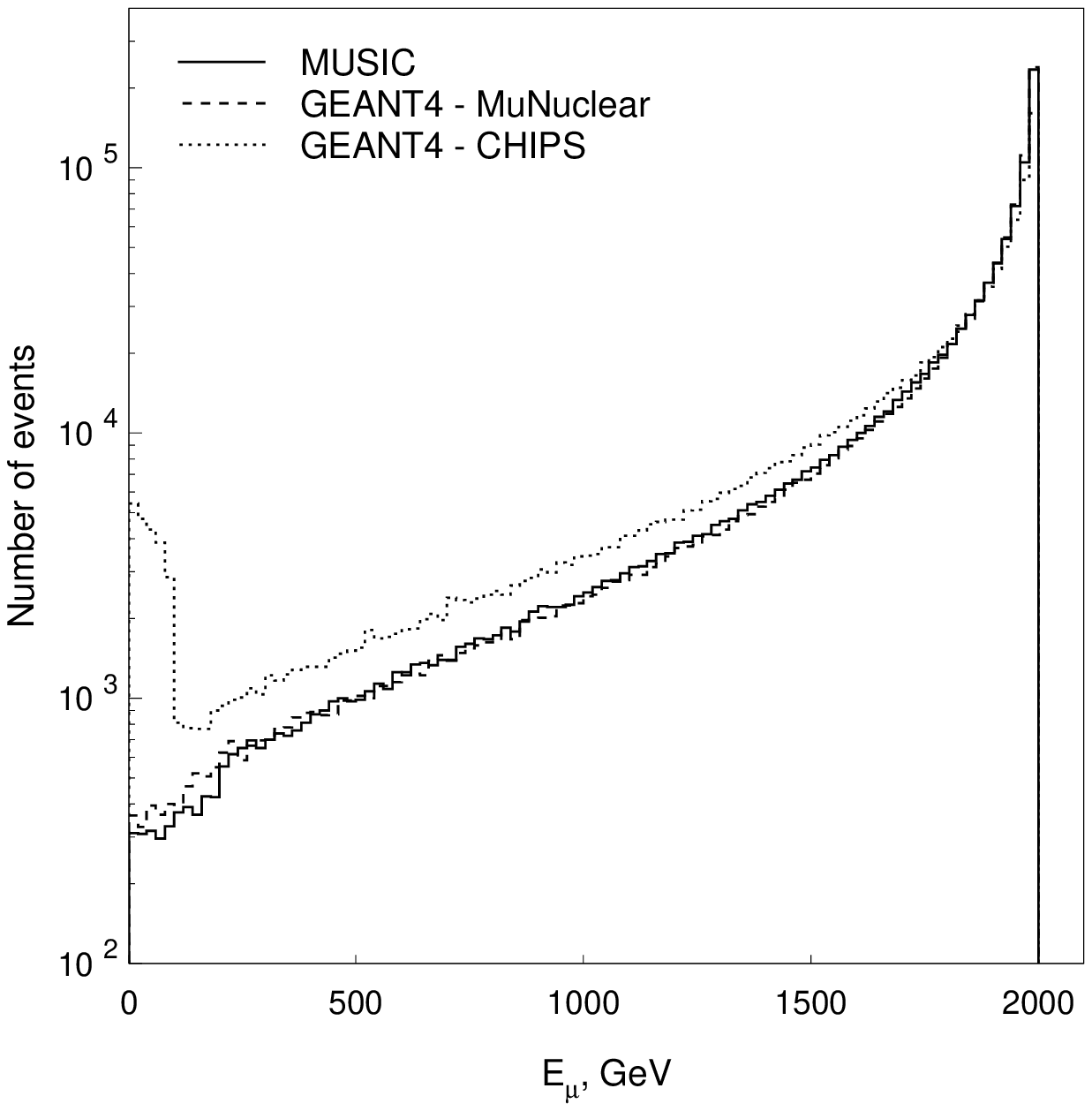}
    \caption{{Energy spectrum of 2 TeV muons after crossing 3 km of water. Only the muon inelastic scattering process was considered. The GEANT4 CHIPS model shows a large increase in the number of muons below 100 GeV, an indication that the model may be unsuited for this energy scale.}}
  \label{fig-transport_3km_water}
\end{center}
\end{figure}

The difference between the MuNuclear and CHIPS models can be seen in Figure \ref{fig-transport_3km_water}, where the energy distributions are plotted after muon propagation taking into account only the muon inelastic scattering process. Muons with 2 TeV energy have been transported through 3 km of water and their energy spectra have been recorded. CHIPS model predicts larger muon energy loss, resulting in the larger number of muons with smaller energies compared to the MuNuclear model or MUSIC. A large enhancement in the number of muons is seen in CHIPS below 100 GeV (5\% of the initial muon energy), an indicator that this process is highly suppressed for low-energy muons in this model.

\section {Simulation of the experiment}
\label{simulation}

\subsection{Experimental setup}
 \label{sim-experiment}
 
The measurement of the muon-induced neutron flux in the Boulby Underground Laboratory (1070~m deep, 2850~w.e. \cite{vitaly08}) was performed using a liquid scintillator detector which served as a veto system for the ZEPLIN-II dark matter detector \cite{z2}. It consists of a large metal container surrounding the lower half of the ZEPLIN-II detector, filled with 0.73 tonnes of liquid scintillator with a known composition and density. It was used to detect both primary high-energy muons (used as trigger) and secondary (low-energy) $\gamma$-rays from neutron capture using delayed coincidences. Independent electronics and data acquisition systems were installed for this measurement, running in parallel with the ZEPLIN-II experiment (see Figure \ref{fig-sim_geometry}). 

ZEPLIN-II and its anti-coincidence system are surrounded by a shielding structure made of lead which was designed to attenuate $\gamma$-rays emitted from the rock walls. This lead `castle' has a thickness of 15~cm on the top section and 22.5~cm on the side walls and floor. Weighing in at approximately 50 tonnes, it creates an excellent target for the production of neutrons by muons. Between the top of the castle and the ZEPLIN-II detector, Gd-impregnated wax (0.2\% Gd by weight) and a thick polypropylene sheet were used for the shielding of rock neutrons. Several layers of polypropylene (interleaved with thin layers of Gd-loaded wax) were also used inside the lead castle, above the veto system. Furthermore, the inner surface of the veto vessel was painted with a mixture containing Gd salt. The purpose of using Gd around the detector was to increase the number of neutron captures, enhancing the probability of neutrons produced by radioactivity in detector components being detected by the veto or the ZEPLIN-II detector itself.

Further details on this experiment, including data acquisition and analysis procedures, can be found in Ref. \cite{vitaly08} and references therein.

\subsection{Simulation details}
 \label{sim-details}
 
The simulation of this experiment can be divided in two stages. In the first, the MUSUN \cite{vak03} code was used to sample muons according to their energy spectrum and angular distribution at the Boulby Underground Laboratory (obtained with the code MUSIC \cite{music} propagating a spectrum of atmospheric muons from the surface through Boulby rock). The muon flux thus obtained is normalised to the one measured in this experiment, which agrees within 10\% with previous measurements \cite{matt03}. This results in a mean energy of $\sim$260 GeV for Boulby underground muons. These primaries were sampled on the surface of a parallelepiped encompassing the experimental hall, with dimensions chosen to give a clearance of 10~m, 7~m and 4~m to the laboratory walls (top, sides and bottom, respectively). Two million of these muons were generated and their properties (energy, momentum, position and charge sign) stored for the second stage.

\begin{figure*}[htb]
\begin{center}
   \includegraphics[width=1.25\columnwidth]{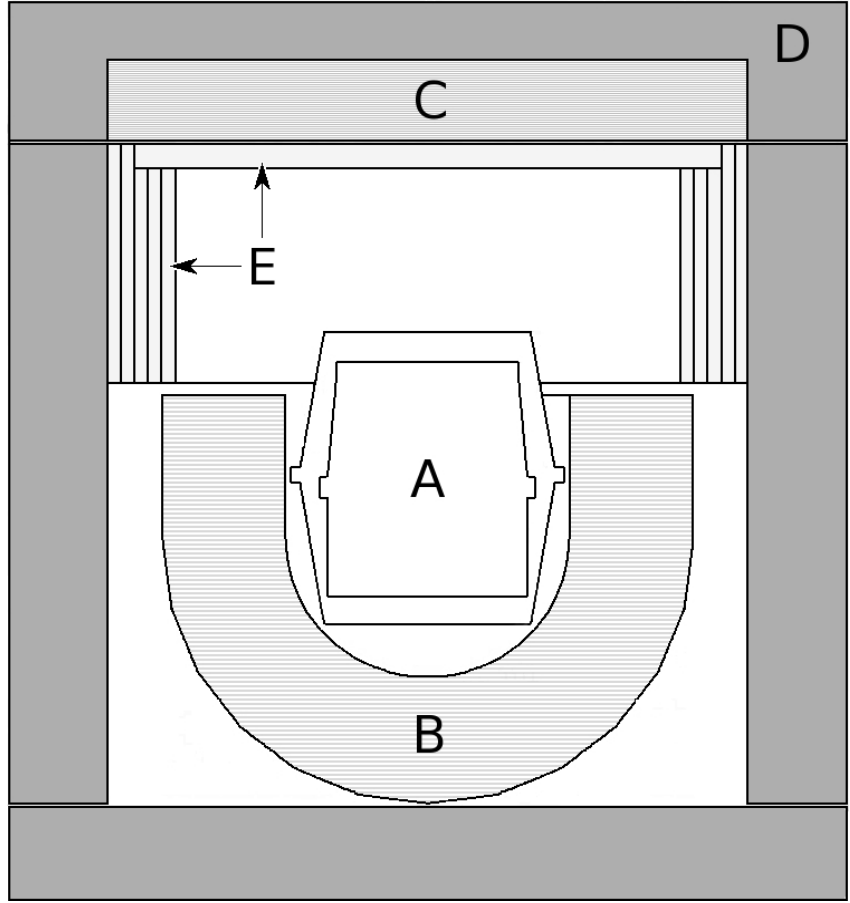}
    \caption{{Vertical cut of the geometry model used in the GEANT4 simulation: A -- ZEPLIN-II detector, B -- liquid scintillator detector (veto), C -- Gd-loaded wax, D -- lead castle, E -- polypropylene sheets which make up the passive neutron shielding (vertical slabs are interleaved with Gd-loaded resin). Details of the ZEPLIN-II detector were removed from this figure for simplicity.}}
  \label{fig-sim_geometry}
\end{center}
\end{figure*}

A detailed GEANT4 simulation of the experiment was developed for the second stage using version 8.2 of this toolkit. The experimental hall, lead castle, veto detector and the entire ZEPLIN-II detector were included, as well as the neutron shield (with Gd-loaded materials). A thin layer of Gd was also applied to the inner surfaces of the veto, in order to mimic the Gd-loaded paint. Figure \ref{fig-sim_geometry} shows a cross-sectional view of the final geometry obtained with GEANT4.

Muons sampled in the first stage are the input to this simulation. The primary muons and all secondary particles created in the muon-induced cascades are propagated inside this geometry using the set of physics models described in Section \ref{models}. Each energy deposit inside the veto is recorded in a time histogram, which can be directly compared with the experimental results. Energy deposits inside the ZEPLIN-II active volume are also recorded to allow coincidence studies. For neutron captures, the capture material and isotope are stored, together with information about the production of the captured neutron: material, parent particle and production process. For the case of inelastically scattered neutrons we postulate that the highest energy neutron in the final state keeps the identity of the incident neutron. Capture times are also stored, so that they can be cross-referenced to energy deposits inside the veto.

About 40\% of the experimental data was recorded while the top section of the lead castle (and the Gd-loaded wax, C in Figure \ref{fig-sim_geometry}) and the horizontal polypropylene sheet on top of the detector were not in place -- these are called `roof-off' runs, as opposed to the usual `roof-on' runs performed with the complete shielding. This modification to the geometry was also included in the simulation for a similar fraction of the processed events. In total some 120 million muons were simulated, corresponding to a total live time of 960 days, about 4.7 times longer than the experimental exposure, with each muon generated in the first stage being transported through the rock and setup about 60 times using different sets of random numbers.

 \subsection{Validation}
  \label{sim-validation}

The detector was calibrated in energy with a $^{60}$Co source (at the beginning and end of the run); its sensitivity to neutron captures was tested with a 0.1 GBq ($\alpha$~activity) Am-Be source at the beginning of the run. Both tests were simulated with GEANT4 using the same geometry and set of physics models described above, with the only difference being the primary particles passed to the simulation: for the energy calibration $\gamma$-rays were emitted with energies from either of the two $^{60}$Co lines (1.173~MeV and 1.333~MeV); for the Am-Be run the SOURCES4A \cite{sources4} code was used to generate the energy spectrum of the initial neutrons.

Comparison of these simulations with experimental data \linebreak shows a very good agreement (particularly in terms of time delay distribution of $\gamma$-rays from neutron captures) \cite{vitaly08}, thus validating the geometry implementation and physics models used. Note, however, that the neutron calibration is not ideal for evaluating the detector response to muon-induced neutrons: Am-Be neutrons have, on average, lower energies and are created at a fixed location (just below the castle roof), whereas neutrons originated by muons have higher energies and can be created anywhere. There are also other differences, due to experimental limitations: the triggers used are necessarily different (neutron-induced proton recoils instead of high-energy deposits from muons) and the fact that the acquisition system was running with almost 100\% dead-time during calibration. All these problems are obviously not particular to this experiment, but rather apply to any similar experiment that uses neutron sources for their calibration.

Nevertheless, this calibration allows us to show that this experiment is sensitive to neutrons and that our simulation is accurate enough to simulate neutron transport and detection in the MeV and sub-MeV energy ranges.

\section{Results}
 \label{results} 

Some results from this simulation were already analysed in Ref. \cite{vitaly08}, where it was used as an aid in the interpretation of experimental results.

Comparison of the simulated muon rate with the experimental value of 52.9~$\pm$~0.5 per day allows the estimation of the muon flux at Boulby as (3.79 $\pm$ 0.04~(stat) $\pm$ 0.11~(syst))~$\times$~10$^{-8}$ cm$^{-2}$s$^{-1}$, with the systematic error coming from an uncertainty in the energy scale, discussed in detail in Ref. \cite{vitaly08}. The vertical depth obtained from this measurement is 2850~$\pm$~20~m w.e., with the assumption of a flat relief on the surface. Detailed descriptions of the procedures to estimate the muon flux and depth can be found in Ref. \cite{matt03}. 

As the light collection in the veto was not simulated, the energy deposition was smeared using a Gaussian distribution with a standard deviation, $\sigma$, given by the equation $\sigma/E = \sqrt{\alpha+\beta/E}$~\cite{birks}. The parameters $\alpha$ and $\beta$ were determined by comparing the smeared spectrum with the measurements, and the result was used for the energy calibration of the experimental data and the definition of the threshold for muons (20~MeV) and delayed $\gamma$-rays (0.55~MeV).

The shape of the time delay distributions (of $\gamma$-rays from neutron captures) for `roof-on' and `roof-off' simulations was compared with experimental data (using the combined data from both experimental runs), showing a good agreement in the interval 40~--~190~$\mu$s. The first 20~$\mu$s were not considered due to the high level of after-pulsing in the experimental data following the muon signal, while the period from 20 to 40~$\mu$s was removed because of the high probability of detecting $\gamma$-rays from neutron captures in gadolinium, which is difficult to simulate accurately due to uncertainties in its distribution in the wax, resin and paint materials used. 

The agreement in the shape of the simulated and measured time delay distributions from 40 -- 190 $\mu$s also enables a direct comparison of the neutron rates (defined as the average number of detected neutrons per detected muon), by counting the $\gamma$-rays from captured neutrons inside this interval in both distributions. The average rate obtained from the experimental data is 0.079~$\pm$~0.003 neutrons/muon (0.084~$\pm$~0.004 neutrons/muon for `roof-on', 0.072~$\pm$~0.005 neutrons/muon for `roof-off'). The average rate from the simulation is 0.143~$\pm$~0.002~$\pm$~0.009 neutrons/muon (0.143~$\pm$~0.002 neutrons/muon for `roof-on',\linebreak 0.143~$\pm$~0.003 neutrons/muon for `roof-off'). The systematic uncertainty comes mainly from the already mentioned uncertainty in the energy scale. There is a factor of 1.8 difference between simulated and experimental values, with GEANT4 predicting a higher neutron rate than measured. Given the agreement in the shape of the time delay distribution in both the Am-Be neutron calibration and the background runs, it is unlikely that this difference results from a problem in the geometry setup of the simulation or the capture and low energy transport models of neutrons (these models were also tested in Ref. \cite{lemrani06} and found to be in good agreement with MCNPX \cite{mcnpx}). Therefore, the most probable explanation for the excess of muon-induced neutrons lies in the GEANT4 models involved in neutron production. 

\begin{figure}[htb]
\begin{center}
   \includegraphics[width=1.0\columnwidth]{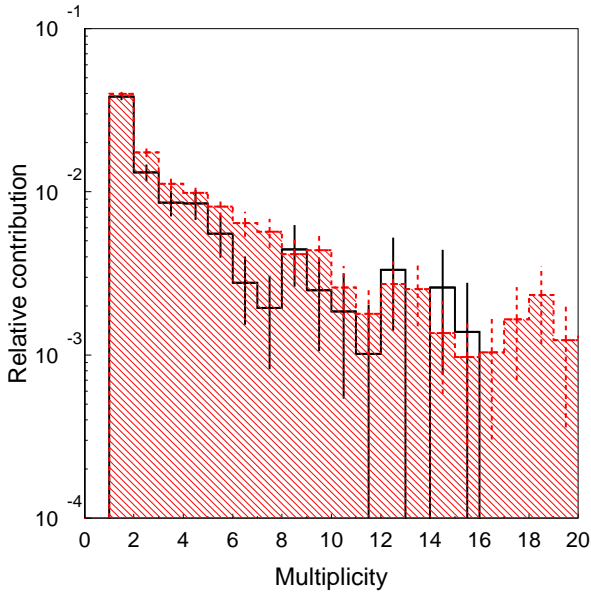}
    \caption{{Relative contribution of each multiplicity to the total number of detected $\gamma$-rays from neutron captures in experimental (solid black) and simulated data (shaded).}}
  \label{fig-mu-multiplicity}
\end{center}
\end{figure}

Figure \ref{fig-mu-multiplicity} shows the relative contribution of each neutron multiplicity to the overall number of detected $\gamma$-rays for both experimental and simulated data. It is clear that GEANT4 produces an excess of neutrons for all multiplicities up to $m=12$. Constraints in the data acquisition system effectively limit the number of recorded pulses to 16, but analysis of simulated events with larger multiplicities show that the fraction of missed neutrons is at the level of a few percent, and thus does not contribute significantly to the observed difference in the rates.

Table \ref{tab-mu-nproduction} shows the contributions of the most relevant materials for neutron production. In the `detected' columns only neutrons with $\gamma$-ray hits above threshold and in the time interval 40~--~190~$\mu$s were used, while `all' includes all neutrons that ended up being captured in the simulation (although very small, there is a finite probability that a neutron leaves the geometrical limits of the simulation without being captured). As expected, nearly all neutrons created in this experiment come from rock, but only fewer than 10\% are detected, attesting the effectiveness of the shield. Lead is by far the biggest producer of detected neutrons, being responsible for $\sim$90\%. Even in `roof-off' runs, when there is a 12\% reduction in Pb mass, its contribution to the number of detected neutrons decreases by less than 1\%. This can be due to two reasons. First, the presence of a large volume of Gd-loaded wax in the roof section enhances the moderation and capture of neutrons in this region, further away from the veto than the H-rich materials. Moreover, the chosen time window deliberately excludes the period in which the captures in Gd dominate. Scintillator itself does not give a noticeable contribution to neutron production.

\begin{table}[htb]
   \centering
   \topcaption{{Contributions of different materials to neutron production for captured neutrons. `Detected' means neutrons which deposited at least 0.55~MeV in the veto (after gaussian smearing) during the interval 40~--~190~$\mu$s in events with a trigger signal (muon energy deposition) of at least 20~MeV; `all' refers to all captured neutrons. `Roof-on' and `roof-off' refer to, respectively, runs with the complete shielding and runs without the roof sections of the Pb castle and neutron passive shielding.}}
   \begin{tabular}{@{} l|cc|c @{}} 
	\hline
      Material	& \multicolumn{2}{c|}{detected} & all\\
      			& roof-on	& roof-off	& \\ 
      \hline
      Lead		& 91.4\%	& 90.6\%	& 0.6\%	\\
      Rock		& 6.1\%	& 8.0\%	& 99.4\%	\\
      Copper	& 2.5\%	& 1.4\%	& --		\\
      \hline
   \end{tabular}
   \label{tab-mu-nproduction}
\end{table}

Lead contributes 90\% to the neutron production, making uncertainties in other materials and elements of the geometry relatively unimportant. The increased neutron rate in the simulation therefore implies that GEANT4 overestimates the neutron production in this material. We note that FLUKA predicts a higher rate, again 80\% in excess of GEANT4. This also contradicts the NA55 results, as discussed in Section \ref{mu_spallation}. Moreover, the alternative (CHIPS-based) models yield a $\sim$1.6$\times$ higher rate, and thus go in the wrong direction regarding these experimental results.

From the simulation we find that only 42\% of neutrons are detected in the time window 40~--~190~$\mu$s, and can thus estimate the experimental rate for an infinite time window to be 0.188~$\pm$~0.005 neutrons/muon. Considering that most neutrons are created in lead and that we have confidence in the simulated geometry and the GEANT4 models that treat neutron moderation and capture, and assuming that the ratio of simulated-to-measured neutron rates is the same as for raw neutron yields in the material, it is possible to estimate the absolute neutron yield of 260 GeV muons in lead. Using the neutron yield obtained in Section \ref{neutron_yield} for 260~GeV neutrons (2.37~$\times~$10$^{-3}$~neu\-trons/muon/(g/cm$^2$)) and the 1.8 ratio between the simulated and measured rates, we estimate an actual neutron yield in lead of (1.31~$\pm$~0.06)~$\times$~10$^{-3}$~neu\-trons/muon/(g/cm$^2$), which is 3.2 times smaller than expected from FLUKA-2008.

In Table \ref{tab-mu-ncapture} the contribution of different elements to the capture of neutrons is shown for the two situations described above (`detected' and `all'). As expected, hydrogen is responsible for almost two thirds of the detected neutrons, again with little difference between the `roof-on' and `roof-off' situations. Scintillator is obviously the main contributor to detected hydrogen captures, with 85\% of the neutrons being captured in this material, while polypropylene is responsible for only 15\%. The small contribution from Gd is again due to the choice of the time interval for detected pulses. Na and Cl (the main constituents of the Boulby rock) are responsible for almost all the neutron captures (99.7\%), but have a small contribution to the detected $\gamma$-rays even when the roof of the lead castle was not in place, an effect of the reduced solid angle and the selection cuts used in the experimental data (threshold and time interval).

\begin{table}[htb]
   \centering
   \topcaption{{Contributions of different elements to neutron captures. `Detected' means neutrons which deposited at least 0.55~MeV in the veto (after gaussian smearing) during the interval 40~$\mu$s to 190~$\mu$s in events with a trigger signal (muon energy deposition) of at least 20~MeV; `all' refers to all captured neutrons. `Roof-on' and `roof-off' refer to, respectively, runs with the complete shielding and runs without the roof sections of the lead castle and neutron passive shielding.}} 
   \begin{tabular}{@{} l|cc|c @{}} 
	\hline
      Element		& \multicolumn{2}{c|}{detected} & all \\		 
      				& roof-on	& roof-off			& \\
      \hline
      H				& 65.7\%	& 64.4\%			& 0.1\%	\\	
      Gd			& 12.5\%	& 10.0\%			& 0.1\%	\\	
      Cl				& 12.1\%	& 14.0\%			& 94.5\%	\\	
      Cu			& 4.0\%	& 4.4\%			& --		\\		
      Fe				& 4.0\%	& 3.4\%			& --		\\		
      Na			& 0.3\%	& 0.5\%			& 5.2\%	\\		
      others		& 1.4\%	& 3.3\%			& 0.1\%	\\
      \hline
   \end{tabular}
   \label{tab-mu-ncapture}
\end{table}

The contribution of muon-induced neutrons to the background of the ZEPLIN-II detector was also estimated using this simulation. For this study 55.5~million muons were simulated (corresponding to 1.23~years exposure) using the `roof-on' geometry, and energy depositions in the xenon were recorded along with the hits in the veto to allow anti-coincidence studies. Moreover, nuclear recoils (NR) were recorded separately and normalised by a conversion factor to obtain the equivalent energy for electromagnetic interactions (E$_{ee}$): $E_{NR} = E_{ee}/0.36$ \cite{z2}.

The results of this study are summarised in Table \ref{tab-zeplin2-nr}, where the contribution of different types of NR events to the background is shown, using thresholds and energy ranges of interest for dark matter experiments (namely the interval 2~--~20~keV where most WIMP interactions are expected to occur). The table lists both `pure' and `mixed' recoil events, the latter representing events that involve both NR and electromagnetic energy depositions.

\begin{table}[htb]
   \centering
   \topcaption{{Muon-induced neutron background (NR events per year) in the ZEPLIN-II (31~kg) target volume for different detection thresholds and energy intervals.}} 
   \begin{tabular}{@{} lcc @{}} 
      \hline
	Event type						& E$_{ee}$ (keV)	& Hits/year\\ 
      \hline
	Mixed NR events					& $>0$				& 57.7 $\pm$ 6.9\\
									&					&	\\
	Pure NR events				& $>0$				& 13 $\pm$ 3.3\\
									& $>2$				& 3.3 $\pm$ 1.6\\
									& 2 -- 20			& 3.3 $\pm$ 1.6\\
									&					&	\\
	Pure single NR events		& $>0$				& 5.7 $\pm$ 2.2\\
									& 2 -- 20			& 0.8 $\pm$ 0.8\\
									&					&	\\
	Anti-coincidence with veto	& $>0$				& $<1.9$ (90\% C.L.)	\\
      \hline
   \end{tabular}
   \label{tab-zeplin2-nr}
\end{table}

The predicted number of events involving at least one nuclear recoil (and with any energy deposition) in the target volume of ZEPLIN-II is 57.7~$\pm$~6.9 per year, of which 13~$\pm$~3.3 have no coincident electromagnetic deposition. 3.3~$\pm$~1.6 of these events are expected to fall in the interesting energy range for WIMP search, but this number can be further reduced to 0.8~$\pm$~0.8 if we consider that the detector has enough spatial resolution to resolve the locations of multiple neutron elastic interactions and remove these events. Nevertheless, when we introduce an anti-coincidence cut with the veto signal (even using a threshold as high as 1~MeV) no NR event survives. If only the inner volume of ZEPLIN-II is used in the data analysis (7.2~kg) these neutron rates are reduced by approximately a factor of 4.

Data from the ZEPLIN-II science run \cite{z2}, with a total exposure of 225~kg$\times$days, was re-analysed to search for neutron recoils in coincidence with a large muon signal in the veto, but none was found.

These results show that the muon-induced background is not a severe threat for the current generation of detectors, with target masses up to a few tens of kilograms, but may become a problem for larger scale targets.

\section {Conclusions}
\label{conclusions}

Version 8.2 of the GEANT4 toolkit was used to create a detailed Monte Carlo simulation of the first experiment to measure the muon-induced neutron background at the Boulby Underground Laboratory, which used the 0.73 tonne liquid scintillator veto from the ZEPLIN-I and ZEPLIN-II dark matter search experiments. 

First, the neutron yields of mono-energetic muons in different materials of interest to this experiment were compared with results obtained with older versions of this toolkit and other simulation packages, to benchmark recent modifications to the relevant models. Yields are consistently lower than the FLUKA predictions for all three materials tested and 280~GeV muons, the differences being 30\% for C$_n$H$_{2n}$, 50\% for NaCl and 80\% for Pb, but are still in good agreement with version 6.2 of GEANT4. Newly available CHIPS-based models were also tested while attempting to describe results from the NA55 experiment, which tried to measure the neutron production in thin targets of different materials. These new models clearly produce higher yields for all materials and detection angles, but are still far from explaining the controversial NA55 results.

A direct comparison of the simulated and measured neutron capture rates in the 40--190~$\mu$s window after the trigger reveals that the simulation predicts 1.8 times more neutrons than measured (well beyond statistical and systematical errors), with a similar contribution from mutiplicities up to 12. Given that Pb is responsible for $\sim$90\% of the detected neutrons, this implies that the GEANT4 models over-predict the neutron production in this material, contrary to what was expected from the comparison with FLUKA predictions and the NA55 results. The neutron yield in lead was estimated to be (1.31~$\pm$~0.06)~$\times$~10$^{-3}$ neutrons/muon/(g/cm$^2$).

Finally, the contribution of muon-induced neutrons to the ZEPLIN-II background was estimated. The expected single nuclear recoil rate is less than one event per year, but an anti-coincidence with the veto effectively removes all nuclear recoil events, confirming that this is not a meaningful source of background for a detector of this size. 

\section{Acknowledgments}

This work has been supported by the ILIAS integrating activity (Contract No. RII3-CT-2004-506222) as part of the EU FP6 programme in Astroparticle Physics. We acknowledge the financial support from the portuguese Funda\c c\~ao para a Ci\^encia e a Tecnologia (through the PhD grant for A. Lindote -- Ref. SFRH/BD/12843/2003 -- and project grant POCI/FP/81928/2007) and the UK Science and Technology Facilities Council (STFC). The authors acknowledge the contribution of C. Bungau, C. Ghag and M. Carson to parts of the geometrical model used in this work.

\pagebreak

\pagebreak

\pagebreak

\end {document}